\documentclass{birkjour_t2}
\usepackage{amsmath,amssymb,amsthm,amsfonts}
\usepackage{mathabx}
\usepackage{mathrsfs}
\usepackage{bm}
\usepackage{graphicx}
\usepackage{float}
\usepackage{subfig}
\usepackage{color,xcolor}
\usepackage{cite}
\usepackage[colorlinks=true,linkcolor=magenta, citecolor=teal]{hyperref}

\usepackage{changes}
\numberwithin{equation}{section}

\newcommand{\be}{\begin{equation}}
\newcommand{\en}{\end{equation}}
\newcommand{\la}{\label}
\newcommand{\ep}{{\varepsilon}}
\newcommand{\paa}{\partial}
\def\rr#1{\eqref{#1}}
\def\ii{{\rm i}}

\newcommand{\B}{\mathcal{B}}
\DeclareMathOperator{\tr}{tr}

\title[An analytic derivation of the bifurcation conditions for localization]{An analytic derivation of the bifurcation conditions for localization in hyperelastic tubes and sheets}

\author{Xiang Yu}
\address{School of Computer Science and Technology, Dongguan University of Technology, Dongguan 523808, China\\ Department of Mechanics, School of Mechanical Engineering, Tianjin University, Tianjin 300354, China }
\email{yuxiang@dgut.edu.cn}

\author{Yibin Fu}
\address{School of Computing and Mathematics, Keele University, Staffordshire ST5 5BG, UK}
\email{y.fu@keele.ac.uk}

\begin{document}

\maketitle

\begin{abstract}
We provide an analytic derivation of the bifurcation conditions for localized bulging in an inflated hyperelastic tube of arbitrary wall thickness  and axisymmetric necking in a hyperelastic sheet under equibiaxial stretching. It has previously been shown numerically that the bifurcation condition for the former problem is equivalent to the vanishing of the Jacobian determinant of the internal pressure $P$ and resultant axial force $N$, with each of them viewed as a function of the azimuthal stretch on the inner surface and the axial stretch. This equivalence is established here analytically. For the latter problem for which it has recently been shown that the bifurcation condition is not given by a Jacobian determinant equal to zero, we explain why this is the case and provide an alternative interpretation.
\vspace{1em}


\vspace{1em}

\end{abstract}

\section{Introduction}

We revisit here the problem of localized bulging in a hyperelastic tube of arbitrary wall thickness subject to axial loading and internal pressure, and the problem of axisymmetric necking in a hyperelastic sheet under equibiaxial stretching. Studies on the former problem date back as early as Mallock \cite{M91}, and much progress has been made since then \cite{Y77,CH,RM,ARM,KY,KYb,PGL,GPL}, but misconceptions also persisted that prevented a thorough understanding of this important prototypical localization problem. For instance, localized bulging of inflated rubber tubes was thought to have some connection with the pressure versus volume curve having a maximum (the so-called limiting point instability) \cite{A,KH,HNH}, but the precise nature of this connection was not clear and the initiation pressure was often incorrectly calculated as the bifurcation pressure for a periodic mode. Fu  {\it et al.} \cite{FPL} demonstrated explicitly, under the membrane assumption, that localized bulging is a bifurcation phenomenon but is not connected with a periodic mode. In fact, a weakly nonlinear analysis based on the periodic mode viewpoint would give a bulging profile that has no resemblance to the actual bulging profile observed experimentally or simulated numerically. They also demonstrated that the initiation pressure is equal to the pressure for the limiting point instability in one loading scenario, but this connection may be lost in other loading scenarios (e.g., the case of fixed ends).

Recent studies have focused on tubes of {\it arbitrary} wall thickness to which the membrane assumption no longer applies. With the help of dynamical systems theory, Fu {\it et al.} \cite{FLF} derived the bifurcation condition for localized bulging and showed that it takes a simple form $J(P,N)=0$ where $J(P,N)$ denotes the Jacobian determinant of the internal pressure $P$ and resultant axial force $N$, each viewed as a function of two principal stretches. This bifurcation condition was rederived in \cite{YLF} as a by-product of a weakly nonlinear analysis to derive the bulging solution explicitly. The derived analytic predictions were  corroborated by numerical simulations \cite{LLY} and experimental studies \cite{WGZLF}.
This bifurcation condition provides a framework under which various other effects may be assessed in a systematic manner \cite{WAF,VD,WF,LYAX,YLAX,HHS}. More recently, the same methodology has been applied to study elasto-capillary bulging and necking in soft elastic cylinders \cite{FJG} and tubes \cite{EFa,EFb}. The work \cite{FJG} seems to be the first self-consistent study on this problem that not only addresses the initial bifurcation, but also connects it to the final Maxwell state, correcting again misconceptions in the relevant field.

The bifurcation condition $J(P,N)=0$ was established in \cite{FLF} via a brute force approach: the condition that zero becomes a triple eigenvalue of the spatial dynamical system governing axisymmetric incremental deformations is first derived, and was then shown {\it numerically} to be equivalent to $J(P,N)=0$. The purpose of the current paper is to derive this equivalence {\it analytically}, thus providing further insight into the bifurcation condition and justifying its application in other elastic localization problems. The main idea is to recognize that the bifurcation condition is simply the solvability condition for an extra uniform expansion to exist.

The same methodology is then applied to derive the bifurcation condition for the axisymmetric necking of a thin sheet that is subject to equibiaxial stretching within the plane. Under general biaxial stretching (not necessarily equibiaxial), the two nominal stresses are functions of the two in-plane stretches, and it is then natural to compute their Jacobian determinant, evaluate it at an equibiaxial stretching state, and ask whether it is related to necking. It turns out that this is not the case \cite{WJF2022}, and we explain why.

The rest of this paper is organized as follows. The next section is devoted to the inflation problem. We first summarize the solution for the primary inflation solution and the incremental boundary value problem, and then re-derive the bifurcation solution using a procedure that is simpler than that employed in \cite{FLF}. This new derivation shows explicitly that the bifurcation condition is in fact the solvability condition for a non-trivial uniform perturbation to exist, and thus enables the above-mentioned equivalence to be established. In Section \ref{sec:necking} we use the same methodology to study the axisymmetric necking problem. The paper is concluded in Section \ref{sec:conclusion} with a summary and a discussion of other applications of the methodology proposed in the current paper.

\section{Localized bulging in an inflated hyperelastic tube}\label{sec:bulging}

\subsection{Uniform inflation and extension}\label{sec:problem}

\begin{figure}[h!]
	\centering  
	\subfloat[]{\includegraphics[width=0.29\textwidth]{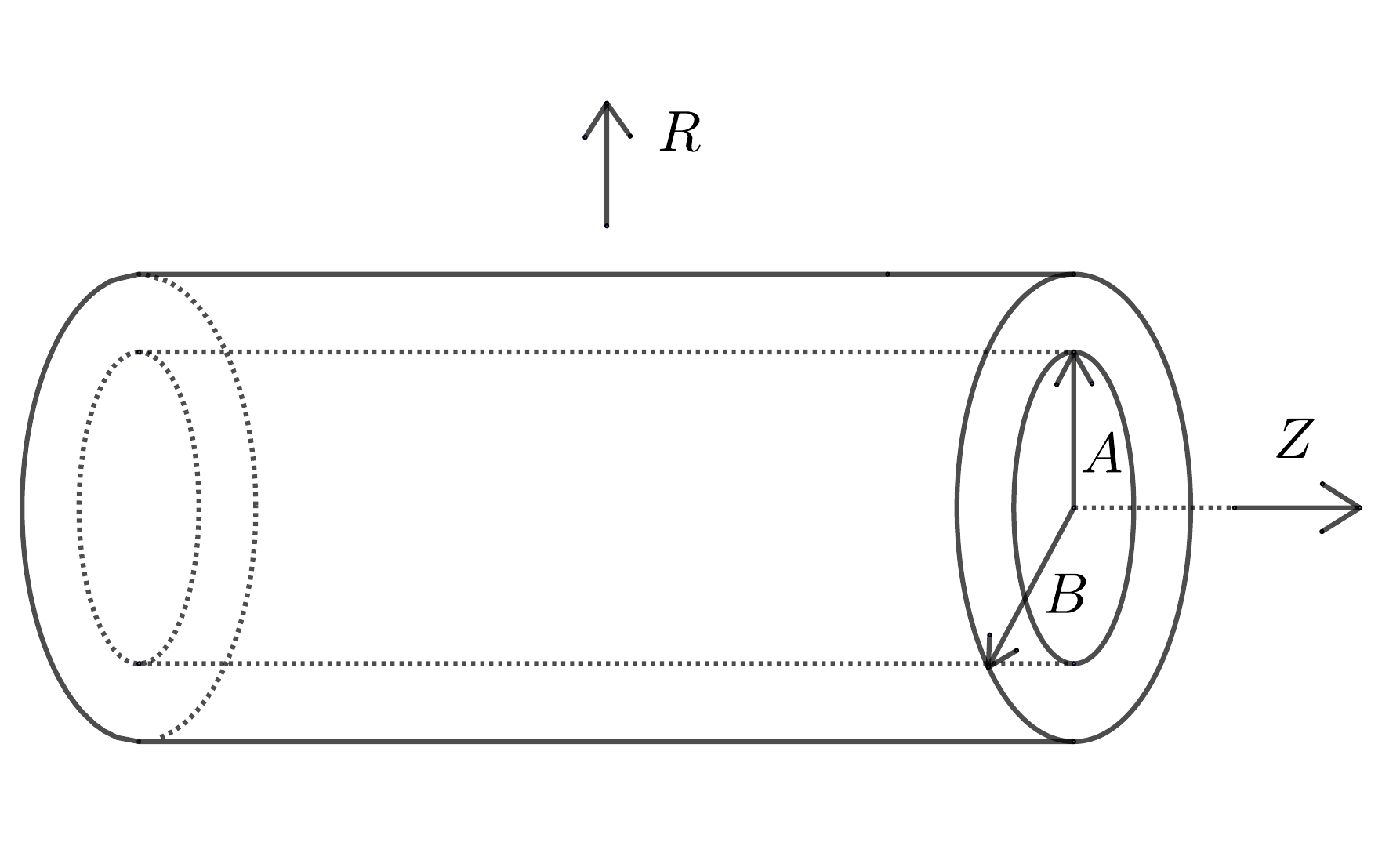}\label{subfig:1a}
	}\qquad
	\subfloat[]{\includegraphics[width=0.41\textwidth]{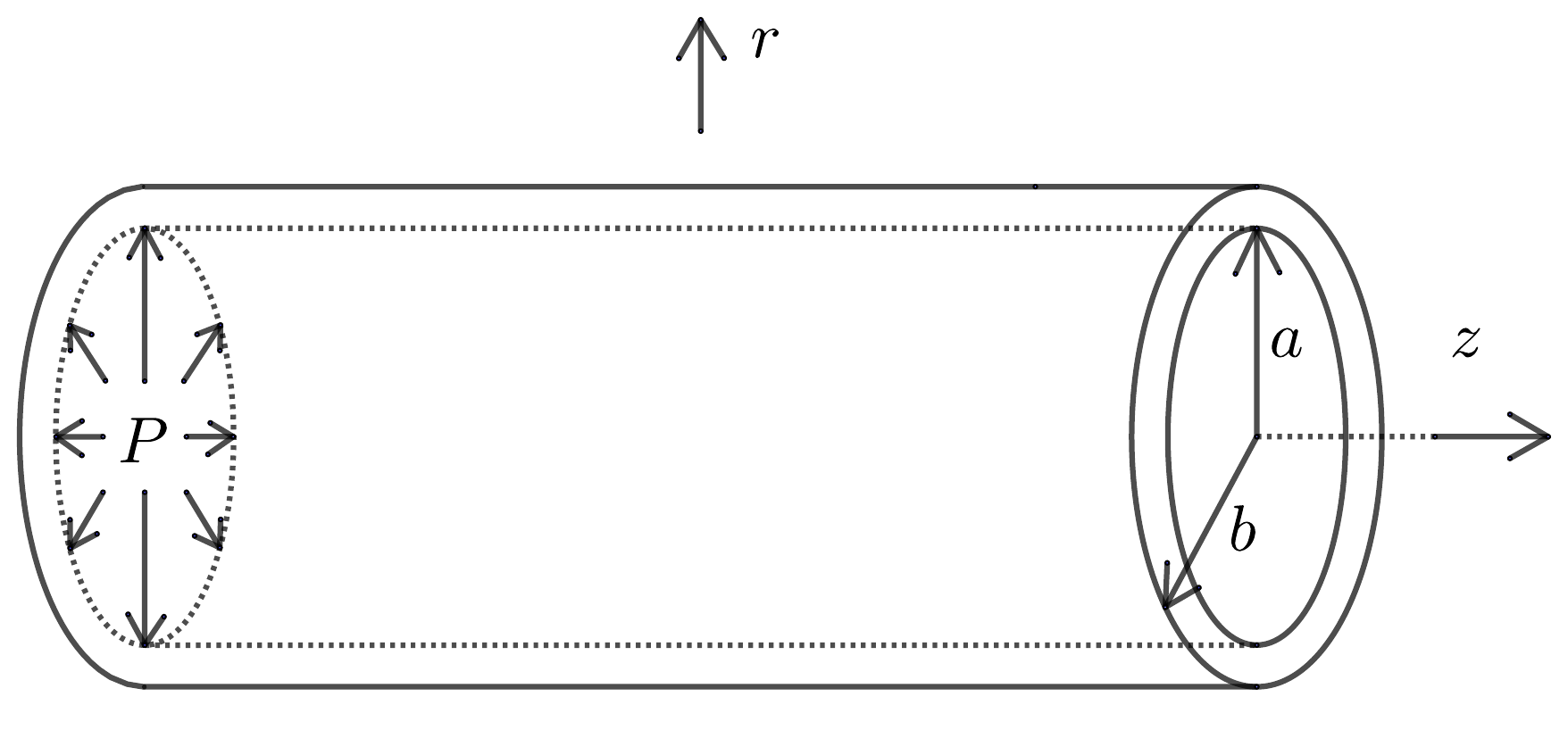}\label{subfig:1b}
	}
\caption{A hyperelastic cylindrical tube in (a) reference (undeformed) configuration and (b) uniformly inflated configuration.}
\label{fig:reun}
\end{figure}

We consider a sufficiently long circular cylindrical tube that is incompressible, isotropic and hyperelastic. The tube is assumed to have inner radius $A$ and outer radius $B$ before deformation; see Fig. \ref{fig:reun}(a). When it is uniformly stretched in the axial direction by a force $N$ and inflated by an internal pressure $P$, the inner and outer radii become $a$ and $b$, respectively, as shown in Fig. \ref{fig:reun}(b). The deformation, in terms of cylindrical polar coordinates, is specified by
\begin{align}\label{eq:rz}
r^2=\lambda_z^{-1}(R^2-A^2)+a^2,\quad \theta=\Theta,\quad z=\lambda_z Z,
\end{align}
where $(R,\Theta, Z) $ and $(r,\theta, z)$ are the coordinates in the undeformed and deformed configurations, respectively,  and $\lambda_z$ is the constant stretch in the axial direction. The first equation in \rr{eq:rz} is a consequence of the incompressibility constraint. It  follows from  \eqref{eq:rz} that the three principal stretches are simply
\begin{align}
\lambda_1=\frac{r}{R},\quad \lambda_2=\lambda_z,\quad \lambda_3=1/(\lambda_1\lambda_2),
\end{align}
where we have identified the {\it  indices $1, 2, 3$ with the $\theta$-, $z$-, and $r$-directions}, respectively. Throughout this paper, we shall refer to the deformed configuration corresponding to \rr{eq:rz} as the {\it uniformly inflated configuration}.

We assume that the constitutive behavior of the tube is described by a strain energy function $W(\lambda_1,\lambda_2,\lambda_3)$. The non-zero Cauchy stresses are given by
\begin{align}\label{eq:22}
\sigma_{ii}=\lambda_i \frac{\partial W}{\partial\lambda_i}-\overline{p},\quad i=1,2,3,
\end{align}
where  $\overline{p}$ is the Lagrange multiplier enforcing the incompressibility constraint. For the considered deformation, $W$ can be regarded as a function of $\lambda_1$ and $\lambda_2$ which we write as $w(\lambda_1,\lambda_2)=W(\lambda_1,\lambda_2,\lambda_1^{-1}\lambda_2^{-1})$. By a standard calculation using \eqref{eq:22} we obtain the stress differences
\begin{align}\label{eq:diff}
\sigma_{11}-\sigma_{33}=\lambda_1w_1,\quad \sigma_{22}-\sigma_{33}=\lambda_2w_2,
\end{align}
where $w_1=\partial w/\partial\lambda_1$ and $w_2=\partial w/\partial w_2$.

The only  equilibrium equation that is not satisfied automatically is
\begin{align}
&\frac{d\sigma_{33}}{dr}+\frac{\sigma_{33}-\sigma_{11}}{r}=\frac{d\sigma_{33}}{dr}-\frac{\lambda_1w_1}{r}=0,\label{eq:radial}
\end{align}
and the associated boundary conditions are
\begin{align}
&\sigma_{33}|_{r=a}=-P,\quad \sigma_{33}|_{r=b}=0, \label{eq:rbc}\\
&\int_a^b r\sigma_{22}\,dr-\frac{1}{2}a^2P =\frac{N}{2\pi}. \label{eq:zbc}
\end{align}
Integrating equation \eqref{eq:radial} subject to the boundary conditions \eqref{eq:rbc}  leads to
\be
-\int_a^b \frac{\lambda w_1}{r}\,dr +P=0, \label{eq:P}
\en
whereas eliminating $\sigma_{22}$ in \eqref{eq:zbc} in favor of $\sigma_{33}$  with the aid of \eqref{eq:diff}$_2$, \eqref{eq:radial} and \eqref{eq:rbc} yields
\be
\int_a^b r \lambda_z w_2\,dr-\frac{1}{2}\int_a^b r\lambda w_1\,dr-\frac{N}{2\pi}=0.\label{eq:N}
\en

Alternatively, the last two equations may be manipulated into the form \cite{HO}
\begin{align}
&P=\int_{\lambda_b}^{\lambda_a}\frac{w_1}{\lambda^2\lambda_z-1}\,d\lambda,\label{eq:Pex}\\
&N=\pi A^2(\lambda_a^2\lambda_z-1)\int_{\lambda_b}^{\lambda_a}\frac{2\lambda_z w_2-\lambda w_1}{(\lambda^2\lambda_z-1)^2}\lambda\,d\lambda,\label{eq:Nex}
\end{align}
where the two limits $\lambda_a$ and $\lambda_b$ are defined by $\lambda_a=a/A$ and $\lambda_b=b/B$, and are related to each other by the incompressibility condition $
(\lambda_b^2\lambda_z-1)B^2= (\lambda_a^2\lambda_z-1)A^2$.

\subsection{Derivation of the bifurcation condition for localized bulging}
We first summarize the linearized incremental equations for the problem formulated in the previous subsection. We consider an axisymmetric perturbation of the form
\begin{align}
\delta\bm{x}= u(r,z)\bm{e}_r+v(r,z)\bm{e}_z,
\end{align}
where $\delta\bm{x}$ stands for the increment of the position vector $\bm{x}$ and $(\bm{e}_r,\bm{e}_\theta,\bm{e}_z)$ denotes the standard orthonormal basis for cylindrical polar coordinates $(r,\theta,z)$. It follows that the incremental deformation gradient $\bm{\eta}$ is
\begin{align}\label{eq:eta}
\bm{\eta}=\frac{u}{r}\bm{e}_{\theta}\otimes\bm{e}_{\theta}+v_z\bm{e}_{z}\otimes \bm{e}_z+v_r\bm{e}_z\otimes \bm{e}_r+u_z\bm{e}_r\otimes \bm{e}_z+u_r\bm{e}_r\otimes \bm{e}_r,
\end{align}
where $v_z:=\partial v/\partial z$, $v_r:=\partial v/\partial r$, etc.

The incremental equilibrium equations that are not satisfied automatically are \cite{YLF}
\begin{align}
&\frac{\partial \chi_{22}}{\partial z}+\frac{\partial \chi_{23}}{\partial r}+\frac{\chi_{23}}{r}=0,\label{eq:l2}\\
&\frac{\partial \chi_{33}}{\partial r}+\frac{\partial \chi_{32}}{\partial z}+\frac{\chi_{33}-\chi_{11}}{r}=0,\label{eq:l3}
\end{align}
where the incremental stress components $\chi_{ij}$ are given by
\begin{align}\label{eq:chi}
\chi_{ij}=\B_{jikl}\eta_{lk}+\overline{p}\eta_{ji}-p^*\delta_{ji}.
\end{align}
In the above expression, $\overline{p}$ and $p^*$ are the Lagrange multipliers associated with the deformation \rr{eq:rz} and the incremental deformation, respectively, and $\B_{ijkl}$ are the instantaneous elastic moduli given by \cite{O}
\begin{align}
\begin{split}
&\B_{iijj}=\lambda_i\lambda_j W_{ij},\\
&\B_{ijij}=\frac{\lambda_i W_i-\lambda_j W_j}{\lambda_i^2-\lambda_j^2}\lambda_i^2,\quad \lambda_i\neq\lambda_j,\\
&\B_{ijji}=\B_{ijij}-\lambda_i W_i,\quad i\neq j,
\end{split}
\end{align}
where $W_i=\partial W/\partial\lambda_i$, $W_{ij}=\partial^2 W/\partial\lambda_i\partial\lambda_j$, etc.

The equilibrium equations \rr{eq:l2} and \rr{eq:l3} are to be solved in conjunction with the incompressibility condition
\begin{align}\label{eq:treta}
\tr(\bm{\eta})=u_r+v_z+\frac{u}{r}=0
\end{align}
subject to the incremental boundary conditions
\begin{align}\label{eq:ibc}
(\bm{\chi}\bm{n}-P\bm{\eta}^T\bm{n})|_{r=a}=0,\quad \bm{\chi}\bm{n}|_{r=b}=0,
\end{align}
where $\bm{n}$ denotes the unit normal to the surface where each of the boundary conditions is imposed. Written out explicitly, the above boundary conditions become
\begin{align}
&v_r+u_z=0,\quad  r=a,b,\label{eq:ibcz}\\
&(\B_{3333}-\B_{2233}+\lambda_3 W_3)u_r+(\B_{1133}-\B_{2233})\frac{u}{r}-p^*=0,\quad  r=a,b.\label{eq:ibcr}
\end{align}

To study the bifurcation of the primary deformation determined previously, we look for an eigensolution of the form
\begin{align}\label{eq:uvp0}
u(r,z)=f(r)e^{\alpha z},\quad v(r,z)=\frac{1}{\alpha} g(r)e^{\alpha z},\quad p^*(r,z)=h(r)e^{\alpha z},
\end{align}
where $\alpha$ is a spectral parameter and the functions $f, g$ and $h$ are to be determined.
On substituting these expressions into \rr{eq:l2}, \rr{eq:l3} and \rr{eq:treta}--\rr{eq:ibcr}, we obtain the differential equations
\begin{align}
\begin{split}\label{eq:g}
& \alpha^2(\B_{2233}+\B_{3223})f'(r)+\alpha^2\frac{1}{r}(r(\B_{3223}'+\overline{p}')+\B_{1122}+\B_{3223})f(r)\\
&+\B_{3232}g''(r)+\frac{1}{r}(\B_{3232}+r\B_{3232}')g'(r)+\alpha^2 \B_{2222}g(r)-\alpha^2 h(r)=0,
\end{split}\\
\begin{split}\label{eq:f}
&\B_{3333}f''(r)+\frac{1}{r}(r(\B'_{3333}+\overline{p}')+\B_{3333})f'(r)+\frac{1}{r^2}(\alpha^2 r^2\B_{2323}+r\B_{1133}'-\B_{1111})f(r)\\
&+(\B_{2233}+\B_{2332})g'(r)+\frac{1}{r}(r\B_{2233}'+\B_{2233}-\B_{1122})g(r)-h'(r)=0,
\end{split}\\
& f'(r)+\frac{f(r)}{r}+g(r)=0 \label{eq:h},
\end{align}
and the associated boundary conditions
\begin{align}
& \alpha^2 f(r)+g'(r)=0, \quad r=a,b,\label{eq:bc2}\\
& (\B_{3333}-\B_{2233}+\lambda_3 W_3)f'(r)+\frac{1}{r}(\B_{1133}-\B_{2233})f(r)-h(r)=0, \ r=a,b. \label{eq:bc3}
\end{align}
In the above equations, $\B_{3223}'=d\B_{3223}/dr$, $\overline{p}'=d\overline{p}/dr$, etc. Solving the above eigenvalue problem using the numerical scheme detailed in  \cite{FLF} or \cite{HO}, we may determine the relationship between $\lambda_a$ and $\alpha^2$. For periodic buckling modes, we replace $\alpha$ by $\ii k$ with $k$ denoting the axial wavenumber. The bifurcation condition for such periodic buckling modes has been computed by Haughton \& Ogden \cite{HO}. Here our attention will be focused on the condition when non-trivial solutions with an infinitesimal $\alpha$ may exist.

We thus assume that $\alpha$ is infinitesimal and write $\varepsilon=\alpha^2$. We aim to determine the corresponding principal stretch $\lambda_a$ for which such a small eigenvalue can exist (the other parameter $\lambda_z$ is either fixed or determined by the condition that $N$ is fixed). Since $\varepsilon$ is small, it is natural to look for a solution of the form
\be
\lambda_a=\lambda_{a\text{cr}}+\varepsilon\lambda_0+ O(\varepsilon^2). \la{dev}
\en
Once we have found this asymptotic expression, it is then clear that $\alpha \to 0$ as $\lambda_a \to \lambda_{a\text{cr}}$. In other words, $\lambda_{a\text{cr}}$ is the value of $\lambda_a$ at which zero becomes a triple eigenvalue and is therefore the critical value for localized bulging to take place \cite{K,M,HI}.

Since the eigenvalue problem \rr{eq:g}--\rr{eq:bc3} contains a small parameter $\ep$, it is appropriate to look for an asymptotic solution of the form
\begin{align}\label{eq:uvp}
\begin{split}
&f(r)=\varepsilon f^{(1)}(r)+\ep^2 f^{(2)}(r)+\cdots,\\
&g(r)=\varepsilon g^{(1)}(r)+\ep^2 g^{(2)}(r)+\cdots,\\
&h(r)=\varepsilon h^{(1)}(r)+\ep^2 h^{(2)}(r)+\cdots,
\end{split}
\end{align}
where the functions on the right-hand sides are to be determined at successive orders.  

On substituting \rr{eq:uvp} into \rr{eq:g}, \rr{eq:h}, \rr{eq:bc2}, and then equating the coefficients of $\varepsilon$,  we obtain	
\begin{align}
&\frac{1}{r}\frac{d}{dr}r\B_{3232}\frac{d}{dr}g^{(1)}=0,\quad g^{(1)}+\frac{1}{r}\frac{d}{dr}rf^{(1)}=0,\\
&\frac{d}{dr}g^{(1)}=0,\quad \ r=a,b.
\end{align}
By straightforward integration, we find that
\begin{align}\label{eq:v1u1}
g^{(1)}=-2 c_1,\quad f^{(1)}=c_1r+\frac{c_2}{r},
\end{align}
where $c_1$ and $c_2$ are arbitrary constants.

The solution for $h^{(1)}(r)$ can be found by considering the coefficients of $\varepsilon$ in the equilibrium  equation \rr{eq:f} and the associated boundary conditions \rr{eq:bc3}, which take the form
\begin{align}
&c_1\omega_1(r)+c_2\omega_2(r)-\frac{d}{dr}h^{(1)}=0, \label{eq:pstarr}\\
&c_1D_1(r)+c_2D_2(r)-h^{(1)}=0,\quad  r=a,b,\label{eq:pstar}
\end{align}
where the functions $\omega_1(r)$, $\omega_2(r)$, $D_1(r)$ and $D_2(r)$ are defined by
\begin{align}
&\omega_1(r)=\B'_{1133}+\B'_{3333}-2\B'_{2233}+\bar{p}'+\frac{1}{r}(2\B_{1122}+\B_{3333}-\B_{1111}-2\B_{2233}),\label{eq:alpha1}\\
&\omega_2(r)=\frac{1}{r^2}(\B'_{1133}-\B'_{3333}-\bar{p}')+\frac{1}{r^3}(\B_{3333}-\B_{1111}),\label{eq:alpha2}\\
&D_1(r)=\B_{1133}+\B_{3333}-2\B_{2233}+\lambda_3 W_3, \label{eq:D1}\\
&D_2(r)=\frac{1}{r^2}(\B_{1133}-\B_{3333}-\lambda_3 W_3). \label{eq:D2}
\end{align}
Integrating \eqref{eq:pstarr} from $r=a$ to $r=b$ and making use of the boundary condition \rr{eq:pstar}, we obtain
\begin{align}\label{eq:m1}
m_{11} c_1+m_{12}c_2=0,
\end{align}
where the coefficients $m_{11}$ and $m_{12}$ are  given by
\begin{align}
&m_{11}=\int_a^b \omega_1(r)\,dr+D_1(a)-D_1(b),\label{eq:m11}\\
&m_{12}=\int_a^b \omega_2(r)\,dr+D_2(a)-D_2(b).\label{eq:m12}
\end{align}
Alternatively, equation \rr{eq:m1} can be obtained by  integrating the equilibrium equation \eqref{eq:l3}  from $r=a$ to $r=b$ and making use of the boundary conditions \eqref{eq:ibc}, that is from
\begin{align}\label{eq:PP}
\int_a^b \frac{\partial \chi_{32}}{\partial z}\,dr+\int_a^b \frac{\chi_{33}-\chi_{11}}{r}\,dr-Pu_r|_{r=a}=0.
\end{align}

A second linear equation for $c_1$ and $c_2$ can be derived from overall equilibrium in the $z$-direction:
\begin{align}\label{eq:conservation}
\int_a^b \chi_{22}r\,dr-P au|_{r=a}=0.
\end{align}
This equation follows from integration of the equilibrium equation  \eqref{eq:l2} multiplied by $r$ from $r=a$ to $r=b$, followed by application of the boundary conditions \eqref{eq:ibc} and the decaying conditions as $z \to \pm \infty$ appropriate for a localized solution. Equating the coefficient of $\varepsilon$ in  the above equation then leads to
\begin{align}\label{eq:m2}
m_{21} c_1 +m_{22}c_2=0,
\end{align}
where the coefficients $m_{21}$ and $m_{22}$ are given by
\begin{align}
&m_{21}=\int_a^b \theta_1(r)r\,dr-\frac{1}{2}\int_a^b \omega_1(r)(b^2-r^2)\,dr-\frac{1}{2}D_1(a)(b^2-a^2)-a^2P,\label{eq:m21}\\
&m_{22}=\int_a^b \theta_2(r)r\,dr-\frac{1}{2}\int_a^b\omega_2(r)(b^2-r^2)\,dr-\frac{1}{2}D_2(a)(b^2-a^2)-P,\label{eq:m22}
\end{align}
with $\theta_1(r)$ and $\theta_2(r)$ defined by
\begin{align}
&\theta_1(r)=\B_{1122}+\B_{2233}-2\B_{2222}-2\overline{p},\label{eq:beta1}\\
&\theta_2(r)=\frac{1}{r^2}(\B_{1122}-\B_{2233}). \label{eq:beta2}
\end{align}	
The existence of a nonzero solution to \eqref{eq:m1} and \eqref{eq:m2} requires that
\begin{align}\label{eq:Omega}
\Omega(\lambda_a,\lambda_z):=m_{11}m_{22}-m_{12}m_{21}=0,
\end{align}
which is an equation that must be satisfied by $\lambda_{a\text{cr}}$, and so is the bifurcation condition for localized bulging. We have verified numerically that \rr{eq:Omega} is equivalent to its counterpart in \cite{FLF}. We note that equations \rr{eq:m1} and \rr{eq:m2} are both obtained at leading order due to the use of two integrals of the incremental equilibrium equations, whereas their counterparts in \cite{FLF} were obtained at a higher order.

\subsection{Equivalence between the bifurcation condition and  $J(P,N)=0$}\label{subsec:JPN}
Although the bifurcation condition \rr{eq:Omega} is simpler than its counterpart in \cite{FLF}, it is still too complicated to give a direct analytical proof of its equivalence to $J(P,N)=0$, where $J(P,N)$ denotes the Jacobian of $P$ and $N$ with each of them viewed as a function of $\lambda_a$ and $\lambda_z$ (see \eqref{eq:Pex} and \eqref{eq:Nex}). Previously, this equivalence was only shown   numerically by verifying that the contour plots of $\Omega(\lambda_a,\lambda_z)=0$ and $J(P,N)=0$ in the $(\lambda_a,\lambda_z)$-plane  always coincide. We now establish this equivalence analytically.

We first note that with the use of  \rr{eq:uvp} and \rr{eq:v1u1}, the solution \rr{eq:uvp0} takes the form
\begin{align}
& u(r,z)=\alpha^2 (c_1 r+\frac{c_2}{r}+O(\alpha^2))e^{\alpha z},\\
&v(r,z)= \alpha (-2 c_1+O(\alpha^2)) e^{\alpha z}.
\end{align}
Thus, to order $\alpha^2$ we have
\begin{align}\label{uvnew}
u(r,z)=\alpha^2 ( c_1 r+\frac{c_2}{r}),\quad v(r,z)=-2\alpha c_1- 2\alpha^2 c_1  z.
\end{align}
Consequently, the coordinates of a representative point in the perturbed configuration are given by
\begin{align}\label{eq:rztilde}
\tilde{r}=r + \alpha^2 (c_1 r+\frac{c_2}{r}),\quad \tilde{z}=z-2\alpha^2 c_1 z-2\alpha c_1.
\end{align}
To interpret these two expressions, we view the $r$ and $z$ given by \rr{eq:rz} as functions of $\lambda_a$ and $\lambda_z$ and differentiate them to obtain
\begin{align}
&dr=\frac{\paa r}{\paa \lambda_a} d \lambda_a+\frac{\paa r}{\paa \lambda_z} d \lambda_z=\frac{A^2 \lambda_a}{r} d \lambda_a- \frac{r^2-A^2 \lambda_a^2}{2 r \lambda_z} d \lambda_z,\label{eq:dr}\\
& dz=\lambda_z^{-1}z d \lambda_z ,\label{eq:dz}
\end{align}
where $dr$ denotes the differential of $r$ etc. It can then be verified that equation \rr{eq:rztilde} may be rewritten as $\tilde{r}=r +dr$, $\tilde{z}=z+dz-2\alpha c_1$ provided
\begin{align}\label{eq:final}
d \lambda_a=\alpha^2(c_1\lambda_a+\frac{c_2}{\lambda_a A^2}),\quad d \lambda_z=-2\alpha^2 \lambda_z c_1.
\end{align}
Thus, the perturbed configuration given by \rr{eq:rztilde} is due to a constant perturbation in $\lambda_a$ and $\lambda_z$. In other words, the solution \rr{uvnew} corresponds to a perturbation that takes the hyperelastic tube from one uniformly inflated configuration to another uniformly inflated configuration, plus a rigid body displacement in the axial direction (represented by the term $-2\alpha c_1 $).
Note that  higher order terms are not relevant to the bifurcation condition since as pointed earlier the latter was derived at leading order. Therefore, {\it the bifurcation condition for localized bulging is simply the condition for an adjacent uniformly inflated configuration to exist}.

Let us denote by  $P^*(\lambda_a,\lambda_z)$ and $N^*(\lambda_a,\lambda_z)$ the two functions resulting from viewing  $P$ and $N$ as functions of $\lambda_a$ and $\lambda_z$ (i.e., the right-hand sides of \eqref{eq:Pex} and \eqref{eq:Nex}). Then uniformly inflated configurations are characterized by the following two equations
\begin{align}\label{eq:QM}
P^*(\lambda_a,\lambda_z)=P,\quad N^*(\lambda_a,\lambda_z)=N.
\end{align}
As argued above, the bifurcation occurs when locally the above relation cannot be inverted uniquely. By the implicit function theorem, this implies that  the Jacobian determinant of the functions $P^*$ and $N^*$ is zero at the bifurcation point. We note that $P^*(\lambda_a,\lambda_z)$ and $N^*(\lambda_a,\lambda_z)$ represent the functional dependence of $P$ and $N$ on $\lambda_a$ and $\lambda_z$, respectively. Hence  the bifurcation condition for localized bulging is equivalent to the vanishing of the Jacobian of $P$ and $N$ with them viewed as functions of $\lambda_a$ and $\lambda_z$.

By carefully analyzing the linearized forms of \eqref{eq:QM} and connecting them with equations \eqref{eq:PP} and \eqref{eq:conservation}, one can establish the equality
\begin{align}\label{eq:proof}
\Omega(\lambda_a,\lambda_z)=-\frac{\lambda_z}{\pi \lambda_a A^2}\Big(\frac{\partial P^*}{\partial \lambda_a}\frac{\partial N^*}{\partial\lambda_z}-\frac{\partial P^*}{\partial \lambda_z}\frac{\partial N^*}{\partial\lambda_a}\Big),
\end{align}
which proves the equivalence between the bifurcation condition  and $J(P,N)=0$ explicitly in view of	 \eqref{eq:Omega}. For interested readers, we present the proof of equality \eqref{eq:proof} in Appendix \ref{app:A}.

\section{Axisymmetric necking in a hyperelastic sheet under equibiaxial stretching}\label{sec:necking}

In this section, we address the bifurcation condition for axisymmetric necking in a hyperelastic sheet under equibiaxial stretching. Unlike the problem studied in the previous section, the governing equations in this problem have variable coefficients and thus do not enjoy translational invariance in the direction of localization. It turns out that the bifurcation condition no longer corresponds to the vanishing of a Jacobian determinant.

\subsection{Homogeneous solution corresponding to equibiaxial tension}

We consider a sufficiently large circular incompressible hyperelastic sheet  that is subject to an equibiaxial tension in its plane. The thicknesses of the sheet in the undeformed and deformed configurations are denoted by $2H$ and $2h$, respectively. The equibiaxial deformation is described by
\begin{align}
r=\lambda R,\quad \theta=\Theta,\quad z=\lambda^{-2}Z,
\end{align}
where $(R,\Theta,Z)$ and $(r,\theta,z)$ are the cylindrical polar coordinates in the undeformed and deformed configurations, respectively, and $\lambda$ is the constant stretch in the plane.  It follows that the three principal stretches are given by
\begin{align}
\lambda_1=\lambda_3=\lambda,\quad \lambda_2=\lambda^{-2},
\end{align}
where $1$, $2$, $3$ still correspond to the $\theta$-, $z$- and $r$-directions, respectively.

In terms of the strain energy function $W(\lambda_1,\lambda_2,\lambda_3)$, the non-zero nominal stress components are given by
\begin{align}
S_{ii}=\frac{\partial W}{\partial\lambda_i}-\overline{p}\lambda_i^{-1},\quad i=1,2,3,
\end{align}
where $\overline{p}$ is the Lagrange multiplier enforcing the incompressibility constraint. The top and bottom surfaces of the sheet are assumed to be traction-free, thus $S_{22}=0$. Eliminating $\overline{p}$ using this condition yields
\begin{align}\label{eq:con}
S_{11}(\lambda_1,\lambda_3)=\frac{\partial w}{\partial \lambda_1}(\lambda_1,\lambda_3),\quad S_{33}(\lambda_1,\lambda_3)=\frac{\partial w}{\partial\lambda_3}(\lambda_1,\lambda_3),
\end{align}
where $w(\lambda_1,\lambda_3)=W(\lambda_1,\lambda_1^{-1}\lambda_3^{-1},\lambda_3)$ is the reduced strain energy function. For the homogeneous solution, we have $2 S_{33}=  dw(\lambda,\lambda)/d\lambda$, which allows one to determine the stretch once the traction at the circular edge is specified.

\subsection{Bifurcation condition for axisymmetric necking}

In a similar fashion to Section \ref{sec:bulging}, the bifurcation condition for axisymmetric necking can be obtained by solving an eigenvalue problem. As in that section, we superpose an axisymmetric perturbation of the form $\delta\bm{x}=u(r,z)\bm{e}_r+v(r,z)\bm{e}_z$ to the homogeneous configuration. The linearized incremental governing equations can be written as
\begin{align}
&\frac{\partial \chi_{22}}{\partial z}+\frac{\partial \chi_{23}}{\partial r}+\frac{\chi_{23}}{r}=0,\label{eq:z}\\
&\frac{\partial \chi_{33}}{\partial r}+\frac{\partial \chi_{32}}{\partial z}+\frac{\chi_{33}-\chi_{11}}{r}=0,\label{eq:r}\\
& \eta_{11}+\eta_{22}+\eta_{33}=0,\label{eq:incom}\\
& \chi_{22}=0,\quad \chi_{32}=0,\quad  z=\pm h,\label{eq:bc}
\end{align}
where the incremental deformation gradient $\bm{\eta}$ and stress tensor $\bm{\chi}$ have been defined in \eqref{eq:eta} and \eqref{eq:chi}, respectively.

We look for an eigensolution of the form
\begin{align}
u(r,z)=\frac{1}{\alpha} f(z)I_1(\alpha r),\quad v(r,z)=g(z)I_0(\alpha r),\quad p^*(r,z)=p(z)I_0(\alpha r),
\end{align}
where $\alpha$ plays the role of the spectral parameter, $I_n(x)$, $n=0,1,\dots$ denote the modified Bessel functions of the first kind that obey the  identities
\begin{align}
I_{n+1}(x)=I_{n-1}(x)-\frac{2n}{x}I_n(x),\quad I'_n(x)=\frac{1}{2}(I_{n-1}(x)+I_{n+1}(x)),\la{iden}
\end{align}
and the functions $f$, $g$ and $p$ are to be determined.
On substituting this solution into \eqref{eq:z}--\eqref{eq:bc} and using \rr{iden} to simplify the results, we obtain the differential equations
\begin{align}
&(\B_{1122}+\B_{3223})f'(z)+\B_{2222}g''(z)+\alpha^2\B_{3232}g(z)-p'(z)=0,\label{eq:add1}\\
&\B_{2323}f''(z)+\alpha^2 \B_{1111}f(z)+\alpha^2(\B_{1122}+\B_{2332})g'(z)-\alpha^2 p(z)=0,\\
&f(z)+g'(z)=0,\label{eq:fg}
\end{align}
and the associated boundary conditions
\begin{align}
&\B_{1122}f(z)+(\B_{2222}+\lambda_2 W_2)g'(z)-p(z)=0,\quad z=\pm h,\\
& f'(z)+\alpha^2 g(z)=0,\quad z=\pm h.\label{eq:add5}
\end{align}
We assume that the bifurcation condition for axisymmetric necking still  corresponds to  when a non-trivial solution with an infinitesimal $\alpha$ exist. To find this condition, we let $\varepsilon=\alpha^2 $ and look for an asymptotic solution of the form
\begin{align}\label{eq:fgh}
\begin{split}
&f(z)=\varepsilon f^{(1)}(z)+\ep^2 f^{(2)}(z)+\cdots,\\
&g(z)=\varepsilon g^{(1)}(z)+\ep^2 g^{(2)}(z)+\cdots,\\
&p(z)=\varepsilon p^{(1)}(z)+\ep^2 p^{(2)}(z)+\cdots.
\end{split}
\end{align}
On substituting \eqref{eq:fgh} into \eqref{eq:add1}--$\eqref{eq:add5}$ and equating the coefficient of $\varepsilon$, we obtain a boundary value problem satisfied by $f^{(1)}$, $g^{(1)}$ and $p^{(1)}$ whose solution  is given by
\begin{align}\label{eq:leading}
f^{(1)}=A,\quad g^{(1)}=-A z+B,\quad p^{(1)}=A(\B_{1122}-\B_{2222}-\lambda_2 W_2),
\end{align}
where $A$ and $B$ are arbitrary constants.

By integrating $r$ times the equilibrium equation \rr{eq:r} across the thickness and making use of the boundary conditions \rr{eq:bc}, we obtain
\begin{align}\label{eq:sector}
r\frac{d \tilde{\chi}_{33}}{d r}+\tilde{\chi}_{33}-\tilde{\chi}_{11}=0,
\end{align}
where the stress resultants $\tilde{\chi}_{11}$ and $\tilde{\chi}_{33}$ are defined by 
\begin{align}
\tilde{\chi}_{11}=\int_{-h}^h \chi_{11}\,dz,\quad \tilde{\chi}_{33}=\int_{-h}^h \chi_{33}\,dz.
\end{align}
%Equation \eqref{eq:sector} can also be derived from integration of \eqref{eq:r} from $z=-h$ to $h$ followed by application of the boundary conditions \eqref{eq:bc}.
Substituting \eqref{eq:leading} into \eqref{eq:sector}, we obtain
\begin{align}\label{eq:biff}
\begin{split}
r\frac{d\tilde{\chi}_{33}}{d r}+\tilde{\chi}_{33}-\tilde{\chi}_{11}&=(\varepsilon(\B_{1111}+\B_{2222}-2\B_{1122}+2\lambda_2 W_2)A+O(\varepsilon^2))2h \alpha r I_1(\alpha r)\\
&=\varepsilon^2 (\B_{1111}+\B_{2222}-2\B_{1122}+2\lambda_2 W_2)h r^2 A+O(\varepsilon^3)=0,
\end{split}
\end{align}
which must hold for arbitrary $\varepsilon$.
It follows that the bifurcation condition for axisymmetric necking is \cite{WJF2022}
\begin{align}
\B_{1111}+\B_{2222}-2\B_{1122}+2\lambda_2 W_2=0. \la{bif10}
\end{align}
We remark that the leading order term on the right-hand side of \rr{eq:biff} is of order $\varepsilon^2$ and the bifurcation condition is obtained by setting this $\varepsilon^2$ term to zero. This is in contrast with the situation in the previous problem where the bifurcation equation was obtained by equating an $O(\varepsilon)$ term to zero.

\subsection{Equivalence between the bifurcation condition and $\partial S_{33}/\partial \lambda_3=0$}

In view of \eqref{eq:fg} and \eqref{eq:leading}, the perturbation solution is of the form
\begin{align}\label{eq:uv}
&u(r,z)=\alpha (A+\alpha^2 C'(z)+O(\alpha^4))I_1(\alpha r),\\
&v(r,z)=\alpha^2(-A z+B-\alpha^2 C(z)+O(\alpha^4))I_0(\alpha r),
\end{align}
where $C(z)=-g^{(2)}(z)$.
Note that we cannot obtain the bifurcation condition by expanding $I_0(\alpha r)$ and $I_1(\alpha r)$ in terms of $\alpha$ and then considering the leading order (i.e., $O(\alpha^2)$) terms of $u(r,z)$ and $v(r,z)$ as in Subsection \ref{subsec:JPN}, since now the bifurcation condition is obtained at $O(\alpha^4)$. To order $\alpha^4$, we have
\begin{align}
&u= \frac{1}{2}\alpha^2 A  r+\frac{1}{16}\alpha^4 A r^3+\frac{1}{2}\alpha^4 r C'(z),\\
&v=\alpha^2(-Az+B)+\frac{1}{4}\alpha^4(-Az+B)r^2-\alpha^4 C(z).
\end{align}
Accordingly, the non-zero components of the incremental deformation gradient to order $\alpha^4$ are
\begin{align}\label{eq:etaa}
\begin{split}
&\eta_{11}=\frac{1}{2}\alpha^2 A  +\frac{1}{16}\alpha^4 A r^2+\frac{1}{2}\alpha^4  C'(z),\\
&\eta_{22}=-\alpha^2 A -\frac{1}{4}\alpha^4 A r^2-\alpha^4 C'(z),\\
&\eta_{23}=\frac{1}{2}\alpha^4 (-Az+B)r,\\
&\eta_{32}=\frac{1}{2}\alpha^4 rC''(z),\\
&\eta_{33}=\frac{1}{2}\alpha^2 A+\frac{3}{16}\alpha^4 A r^2+\frac{1}{2}\alpha^4 C'(z).
\end{split}
\end{align}
Let $\bm{F}$ denote the deformation gradient related to the perturbed deformation. It follows from the chain rule that
\begin{align}
\bm{F}=\begin{pmatrix}
\lambda(1+\eta_{11}) & 0 & 0\\
0 & \lambda^{-2}(1+\eta_{22}) & \lambda \eta_{23}\\
0 & \lambda^{-2}\eta_{32} & \lambda(1+\eta_{33})
\end{pmatrix}.
\end{align}
The corresponding principal stretches are then given by
\begin{align}\label{eq:lambda123}
\tilde{\lambda}_1=\lambda(1+\eta_{11}),\quad \tilde{\lambda}_2=\lambda^{-2}(1+\eta_{22})+O(\alpha^8),\quad \tilde{\lambda}_3=\lambda(1+\eta_{33})+O(\alpha^8).
\end{align}
Thus even to order $\alpha^4$, the $\theta$-, $z$- and $r$-directions are still principal directions, and  the constitutive relation \eqref{eq:con} still holds for the perturbed configuration. We note, however, that the deformation is homogeneous (an equibiaxial extension) to order $\alpha^2$, but is non-homogeneous when expanded to order $\alpha^4$.

From the definition of the stress tensor $\bm{\chi}$, it is not hard to see that  \eqref{eq:sector} corresponds to the linearized form of
\begin{align}\label{eq:linearized}
R\frac{d \tilde{S}_{33}}{d R}+\tilde{S}_{33}-\tilde{S}_{11}=0,
\end{align}
where $\tilde{S}_{11}$ and $\tilde{S}_{33}$ are defined by
\begin{align}
\tilde{S}_{11}=\int_{-H}^H S_{11}\,dZ,\quad \tilde{S}_{33}=\int_{-H}^H S_{33}\,dZ.
\end{align}
With the use of \eqref{eq:etaa} and \eqref{eq:lambda123}, we may expand $S_{11}$ as
\begin{align}\label{nov11}
\begin{split}
S_{11}=&S_{11}|_{\lambda_1=\lambda_3=\lambda}+\frac{\paa S_{11}}{\paa \lambda_1} \lambda \eta_{11} +\frac{\paa S_{11}}{\paa \lambda_3} \lambda \eta_{33} +\frac{1}{2} \frac{\paa^2 S_{11}}{\paa \lambda_1^2} (\lambda \eta_{11})^2 \\
&+\frac{\paa^2 S_{11}}{\paa \lambda_1 \lambda_3} \lambda^2 \eta_{11} \eta_{33} +\frac{1}{2} \frac{\paa^2 S_{11}}{\paa \lambda_3^2} (\lambda \eta_{33})^2+\cdots,
\end{split}
\end{align}
where all the partial derivatives are evaluated at $\lambda_1=\lambda_3=\lambda$. A similar expression for $S_{33}$ can be written. On integrating these two expressions from $Z=-H$ to $Z=H$ and then substituting the resulting expressions into \eqref{eq:linearized}, we find
\begin{align}\label{eq:biff2}
\frac{\partial S_{33}}{\partial\lambda_3}\lambda A H  r^2 \alpha^4 +O(\alpha^6)=0.
\end{align}
We observe that the leading order term on the left-hand side is of order $\alpha^4$ and is due to the $r$-dependent terms in $\eta_{11}$ and $\eta_{33}$. Therefore, the bifurcation condition for axisymmetric necking can  be expressed in the simple form
\begin{align}
\frac{\partial S_{33}}{\partial\lambda_3}=0.
\end{align}
Note that it also follows from the definition of $\bm{\chi}$ that \eqref{eq:sector} differs from the linearized form  of \eqref{eq:linearized} by a factor of $\lambda\frac{h}{H}$. A comparison of \eqref{eq:biff} and \eqref{eq:biff2} yields
\begin{align}
\lambda^2\frac{\partial S_{33}}{\partial\lambda_3}=\B_{1111}+\B_{2222}-2\B_{1122}+2\lambda_2 W_2.
\end{align}
This connection was derived in \cite{WJF2022} by expressing both sides in terms of derivatives of the strain energy function, whereas here it is derived without using these expressions explicitly.

Equation \eqref{eq:linearized} represents equilibrium in the $r$-direction of an infinitesimal annular sector. It is obvious that any homogeneous solution in the form of equibiaxial extension would always satisfy this equation. According to the above analysis, the bifurcation condition for axisymmetric necking corresponds to when this equilibrium equation admits a non-homogeneous solution. This explains why the bifurcation condition for necking is not given by a Jacobian determinant equal to zero.

\section{Conclusion}\label{sec:conclusion}

It was shown numerically in \cite{FLF} that the bifurcation condition for localized bulging of an inflated rubber tube is equivalent to $J(P,N)=0$ with $J(P,N)$ denoting the Jacobian determinant of the internal pressure $P$ and resultant axial force $N$ which are viewed as functions of two deformation parameters $\lambda_a$ and $\lambda_z$.  In this paper, we derived this equivalence analytically by employing two integrals of the equilibrium equations together with the observation that the bifurcation condition is the solvability condition for a non-trivial uniform perturbation to exist. The same strategy is applied to the axisymmetric necking of a stretched elastic plate for which it has recently been shown that the bifurcation condition is not given by a Jacobian determinant equal to zero although the perturbation still represents a homogeneous equibiaxial extension to leading order. We give an explanation for the latter fact by deriving the bifurcation condition analytically, and showing that it is the condition for an infinitesimal sectorial element to admit an adjacent non-homogeneous solution. We emphasize that, contrary to the common belief, a homogeneous perturbation does not necessarily imply that the bifurcation condition corresponds to the vanishing of some Jacobian determinant, as shown in the problem of axisymmetric necking. The bifurcation condition should be derived by carefully analyzing the incremental equations for the homogeneous perturbation and connecting them with  the equilibrium equations of the primary deformation; the latter step usually leads to a simple form of the bifurcation condition. We believe that this method can be applied in other similar localization problems such as elasto-capillary necking/bulging in  soft cylinders \cite{FJG} or tubes \cite{EFa,EFb}, necking in solid cylinders under axial stretching \cite{WF2020}. It is also expected that the current methodology can be used to significantly simplify the weakly nonlinear analysis that determines the localized solutions explicitly.

\section*{Acknowledgments}

This work was supported by the National Natural Science Foundation of China (Grant No. 12072224).

\appendix

\section{Proof of equality \eqref{eq:proof}}\label{app:A}

To prove \eqref{eq:proof}, we start by analyzing the linearized forms of \eqref{eq:QM}. Assuming that the perturbed configuration is also uniform, we show below that the linearized forms of the following two equations 
\begin{align}
&-P^*(\lambda_a,\lambda_z)+P=0,\label{eq:Pp}\\
&\frac{1}{2}b^2 (P^*(\lambda_a,\lambda_z)-P)+\frac{1}{2\pi}(N^*(\lambda_a,\lambda_z)-N)=0 \label{eq:Nn}
\end{align}
agree with \eqref{eq:PP} and \eqref{eq:conservation}.

Let us denote by $\overline{\bm{\sigma}}$ and $\bm{\sigma}$ the Cauchy stresses associated with the uniformly inflated configuration and perturbed configuration, respectively. Then it follows from the definition of the incremental stress tensor $\bm{\chi}$ that
\begin{align}\label{eq:sigma}
\bm{\sigma}=(\bm{I}+\bm{\eta})(\overline{\bm{\sigma}}+\bm{\chi}).
\end{align}
Note that the incremental deformation gradient $\bm{\eta}$ is diagonal since the perturbed configuration is uniform. 

Specifying \eqref{eq:Pp} to the perturbed configuration, we see from the definition of $P^*$ that the resulted equation takes the form
\begin{align}\label{eq:ra}
\int_{\tilde{a}}^{\tilde{b}}\frac{\sigma_{33}-\sigma_{11}}{\tilde{r}}\,d\tilde{r}+P=0.
\end{align}
where $\tilde{r}=r+u$ denotes the radius of the tube after perturbation, and $\tilde{a}=\tilde{r}(A)$ and $\tilde{b}=\tilde{r}(B)$. Substituting \eqref{eq:sigma} into \eqref{eq:ra} and making a change of variables (integration by substitution) by applying the incompressibility equality
\begin{align}\label{eq:rtilde}
\tilde{\lambda}_z(\tilde{r}^2-\tilde{a}^2)=\lambda_z (r^2-a^2),
\end{align}
where $\tilde{\lambda}_z$ is the axial stretch of the tube in the perturbed configuration, we obtain
\begin{align}\label{eq:A3}
\int_{a}^b \frac{(1+\eta_{33})(\overline{\sigma}_{33}+\chi_{33})-(1+\eta_{11})(\overline{\sigma}_{11}+\chi_{11})}{r}\frac{\lambda_z r^2}{\tilde{\lambda}_z \tilde{r}^2}\,dr+P=0.
\end{align}
When expanded to linear order, we have
\begin{align}
\frac{\lambda_z r^2}{\tilde{\lambda}_z \tilde{r}^2}=1-2\frac{\tilde{r}-r}{r}-\frac{\tilde{\lambda}_z-\lambda_z}{\lambda_z}=1-2\eta_{11}-\eta_{22}=1-\eta_{11}+\eta_{33},
\end{align}
where we have used incompressibility constraint
$
\eta_{11}+\eta_{22}+\eta_{33}=0
$.
Thus by ignoring nonlinear terms, one can simplify \eqref{eq:A3} as
\begin{align}\label{eq:A6}
\int_a^b \frac{\chi_{33}-\chi_{11}}{r}\,dr+\int_a^b (\overline{\sigma}_{33}\frac{\eta_{33}-\eta_{11}}{r}+ \frac{\overline{\sigma}_{33}-\overline{\sigma}_{11}}{r}\eta_{33})\,dr=0.
\end{align}
The radial equilibrium for the unperturbed deformation implies that
\begin{align}
\frac{d\overline{\sigma}_{33}}{dr}+\frac{\overline{\sigma}_{33}-\overline{\sigma}_{11}}{r}=0.
\end{align}
Using integration by parts, one can write the second integral in \eqref{eq:A6} as
\begin{align}
\begin{split}\label{eq:A9}
\int_a^b (\overline{\sigma}_{33}\frac{\eta_{33}-\eta_{11}}{r}+ \frac{\overline{\sigma}_{33}-\overline{\sigma}_{11}}{r}\eta_{33})\,dr
&=\int_a^b (\overline{\sigma}_{33}\frac{\eta_{33}-\eta_{11}}{r}-\frac{d\overline{\sigma}_{33}}{dr}\eta_{33})\,dr\\
&=-\overline{\sigma}_{33}\eta_{33}|_{r=a}^{r=b}+\int_a^b \overline{\sigma}_{33}(\frac{\eta_{33}-\eta_{11}}{r}+\frac{d\eta_{33}}{dr})\,dr\\
&=-Pu_{r}|_{r=a}+\int_a^b \overline{\sigma}_{33}(\frac{\eta_{33}-\eta_{11}}{r}+\frac{d\eta_{33}}{dr})\,dr,
\end{split}
\end{align}
where use has been made of the boundary conditions $\overline{\sigma}_{33}|_{r=a}=-P$ and $\overline{\sigma}_{33}|_{r=b}=0$.  In view of incompressibility constraint and the fact that $\eta_{22}$ is constant since the perturbed configuration is uniform, we deduce that
\begin{align}
\begin{split}
\frac{\eta_{33}-\eta_{11}}{r}+\frac{d\eta_{33}}{dr}&=\frac{\eta_{33}-\eta_{11}}{r}-\frac{d\eta_{11}}{dr}=\frac{u_r-u/r}{r}-\frac{d}{dr}(\frac{u}{r})=0.
\end{split}
\end{align}
Putting these together,  we see that the linearized form of \eqref{eq:Pp} can be written as
\begin{align}
\int_a^b \frac{\chi_{33}-\chi_{11}}{r}\,dr-Pu_r|_{r=a}=0,
\end{align}
which agrees with \eqref{eq:PP} (note that $\chi_{32}=0$ for uniform inflations).

Equation \eqref{eq:Nn} applied to the perturbed configuration can be written as
\begin{align}
\int_{\tilde{a}}^{\tilde{b}}\sigma_{22}\tilde{r}\,d\tilde{r}-\frac{1}{2}\tilde{a}^2P-\frac{N}{2\pi}=0.
\end{align}
Using \eqref{eq:sigma} and \eqref{eq:rtilde}, we can rewrite the above equation as
\begin{align}
\int_a^b (\overline{\sigma}_{22}+\chi_{22})r\,dr-\frac{1}{2}\tilde{a}^2P-\frac{N}{2\pi}=0.
\end{align}
Its linearized form is
\begin{align}
\int_a^b \chi_{22}r\,dr-Pau|_{r=a}=0,
\end{align}
which is the same as \eqref{eq:conservation}.

Now let $\tilde{\lambda}_a$ and $\tilde{\lambda}_z$ be the two principal stretches of the perturbed configuration, thus
\begin{align}
&\tilde{\lambda}_a=\lambda_a+d\lambda_a=\lambda_a+\alpha^2(c_1\lambda_a+\frac{c_2}{\lambda_a A^2}),\label{eq:A16}\\
&\tilde{\lambda}_z=\lambda_z+d\lambda_z=\lambda_z-2\alpha^2 \lambda_z c_1.\label{eq:A17}
\end{align}
 Then equation \eqref{eq:Pp} applied to the unperturbed and perturbed configurations takes the form
$
-P^*(\lambda_a,\lambda_z)+P=0$ and $-P^*(\tilde{\lambda}_a,\tilde{\lambda}_z)+P=0$, respectively.
Subtraction of these two equalities yields
\begin{align}\label{eq:mn1}
-P^*(\tilde{\lambda}_a,\tilde{\lambda}_z)+P^*(\lambda_a,\lambda_z)=0.
\end{align}
In a similar way, we can deduce from  \eqref{eq:Nn}  that
\begin{align}\label{eq:mn2}
\frac{1}{2}\tilde{b}^2(P^*(\tilde{\lambda}_a,\tilde{\lambda}_z)-P^*(\lambda_a,\lambda_z))+\frac{1}{2\pi}(N^*(\tilde{\lambda}_a,\tilde{\lambda}_z)-N^*(\lambda_a,\lambda_z))=0,
\end{align}
where $\tilde{b}=\tilde{r}(B)$ is the outer radius of the tube after perturbation. Linearizing the above two equations at $(\lambda_a,\lambda_z)$ followed by the use of \eqref{eq:A16} and \eqref{eq:A17}, we obtain
\begin{align}
&(2\lambda_z\frac{\partial P^*}{\partial\lambda_z}-\lambda_a\frac{\partial P^*}{\partial\lambda_a})c_1-\frac{1}{\lambda_aA^2}\frac{\partial P^*}{\partial\lambda_a}c_2=0,\label{eq:mm1}\\
&\Big(\frac{\lambda_a}{2\pi}\frac{\partial N^*}{\partial\lambda_a}+\frac{\lambda_ab^2}{2}\frac{\partial P^*}{\partial\lambda_a}-\frac{\lambda_z}{\pi}\frac{\partial N^*}{\partial \lambda_z}-\lambda_zb^2\frac{\partial P^*}{\partial\lambda_z}\Big)c_1+\Big(\frac{1}{2\pi \lambda_a A^2}\frac{\partial N^*}{\partial\lambda_a}+\frac{b^2}{2\lambda_a A^2}\frac{\partial P^*}{\partial\lambda_a}\Big)c_2=0,\label{eq:mm2}
\end{align}
Comparing them with \eqref{eq:m1} and \eqref{eq:m2}, we conclude that
\begin{align}\label{eq:mm}
\begin{split}
&m_{11}=2\lambda_z\frac{\partial P^*}{\partial\lambda_z}-\lambda_a\frac{\partial P^*}{\partial\lambda_a},\\
&m_{12}=-\frac{1}{\lambda_aA^2}\frac{\partial P^*}{\partial\lambda_a},\\
&m_{21}=\frac{\lambda_a}{2\pi}\frac{\partial N^*}{\partial\lambda_a}+\frac{\lambda_ab^2}{2}\frac{\partial P^*}{\partial\lambda_a}-\frac{\lambda_z}{\pi}\frac{\partial N^*}{\partial \lambda_z}-\lambda_zb^2\frac{\partial P^*}{\partial\lambda_z},\\
&m_{22}=\frac{1}{2\pi \lambda_a A^2}\frac{\partial N^*}{\partial\lambda_a}+\frac{b^2}{2\lambda_a
	A^2}\frac{\partial P^*}{\partial\lambda_a}.
\end{split}
\end{align}
In view of \eqref{eq:Omega}, it follows from \eqref{eq:mm} that $\Omega(\lambda_a,\lambda_z)$ can be expressed as
\begin{align}\label{eq:OmegaJ}
\Omega(\lambda_a,\lambda_z)=-\frac{\lambda_z}{\pi \lambda_a A^2}(\frac{\partial P^*}{\partial \lambda_a}\frac{\partial N^*}{\partial\lambda_z}-\frac{\partial P^*}{\partial \lambda_z}\frac{\partial N^*}{\partial\lambda_a}),
\end{align}
which completes the proof.


\begin{thebibliography}{50}
	
	
	
    \bibitem{M91} Mallock, A.: Note on the instability of India-rubber tubes and balloons when distended by fluid pressure. {\it Proc. Roy. Soc. A} 49, 458--463 (1891). \url{https://doi.org/10.1098/rspl.1890.0116}
	
	\bibitem{Y77} Yin, W.-L.: Non-uniform inflation of a cylindrical elastic membrane and direct determination of the strain energy function. {\it J. Elast.} 7, 265--282 (1977). \url{https://doi.org/10.1007/BF00041073}
	
	\bibitem{CH} Chater, E., Hutchinson, J.W.: On the propagation of bulges and buckles.  {\it J. Appl. Mech.} 51, 269--277 (1984). \url{https://doi.org/10.1115/1.3167611}
	
	\bibitem{KY} Kyriakides, S., Chang, Y.-C.: On the inflation of a long elastic tube in the presence of axial load. Int. J. Solids Struct. 26(9--10), 975--991 (1990). \url{https://doi.org/10.1016/0020-7683(90)90012-K}
	
	\bibitem{KYb} Kyriakides, S.,  Chang, Y.-C.: The initiation and propagation of a localized instability in an inflated elastic tube. {\it Int. J. Solids Struct.} 27(9), 1085--1111 (1991). \url{https://doi.org/10.1016/0020-7683(91)90113-T}

	\bibitem{PGL} Pamplona, D. C., Goncalves, P. B.,  Lopes, S. R. X.: Finite deformations of cylindrical membrane under internal pressure. {\it Int. J. Mech. Sci.} 48(6), 683--696 (2006). \url{https://doi.org/10.1016/j.ijmecsci.2005.12.007}

    \bibitem{GPL} Goncalves, P. B., Pamplona, D.,  Lopes, S. R. X.: Finite deformations of an initially stressed cylindrical shell under internal pressure. {\it Int. J. Mech. Sci.} 50(1), 92--103 (2008). \url{https://doi.org/10.1016/j.ijmecsci.2007.05.001}

	\bibitem{RM} Rodríguez, J.,  Merodio, J.: A new derivation of the bifurcation conditions of inflated cylindrical membranes of elastic material under axial loading. Application to aneurysm formation. {\it Mech. Re. Commun}. 38(3), 203--210, (2011). \url{https://doi.org/10.1016/j.mechrescom.2011.02.004}
	
	\bibitem{ARM} Alhayani, A.A., Rodr\'{i}guez, J., Merodio, J.: Competition between radial expansion and axial propagation in bulging of inflated cylinders with application to aneurysms propagation in arterial wall tissue. {\it Int. J. Eng. Sci.} 85, 74--89 (2014). \url{https://doi.org/10.1016/j.ijengsci.2014.08.008}
	

	

	
	
		 
	\bibitem{A} Alexander, H.: Tensile instability of initially spherical balloons. {\it Int. J. Eng. Sci.} 9, 151--160 (1971). \url{https://doi.org/10.1016/0020-7225(71)90017-6}
	
	\bibitem{KH} Kanner, L. M., Horgan, C.O.: Elastic instabilities for strain-stiffening rubber-like spherical and cylindrical thin shells under inflation. {\it Int. J. Non-linear Mech.} 42, 204--215 (2007). \url{https://doi.org/10.1016/j.ijnonlinmec.2006.10.010}
	
	\bibitem{HNH} Horny, L., Netusil, M., Horak, Z.: Limit point instability in pressurization of anisotropic finitely extensible hyperelastic thin-walled tube. {\it Int. J. Non-linear Mech.} 77, 107--114 (2015). \url{https://doi.org/10.1016/j.ijnonlinmec.2015.08.003}
	
	\bibitem{FPL} Fu, Y.B., Pearce, S.P.,  Liu, K.K.: Post-bifurcation analysis of a thin-walled hyperelastic tube under inflation. {\it Int. J. Non-linear Mech.} 43(8), 697--706 (2008). \url{https://doi.org/10.1016/j.ijnonlinmec.2008.03.003}
	
	
	\bibitem{FLF} Fu, Y.B., Liu, J.L.,  Francisco, G.S.:  Localized bulging in an inflated cylindrical tube of arbitrary thickness--the effect of bending stiffness. {\it J. Mech. Phys. Solids}, 90, 45--60 (2016). \url{https://doi.org/10.1016/j.jmps.2016.02.027}
	
	\bibitem{YLF} Ye, Y., Liu, Y., Fu, Y.B.: Weakly nonlinear analysis of localized bulging of an inflated hyperelastic tube of arbitrary wall thickness. {\it J. Mech. Phys. Solids} 135, 103804 (2020). \url{https://doi.org/10.1016/j.jmps.2019.103804}
	
	\bibitem{LLY} Lin, Z., Li, L.,  Ye, Y.: Numerical simulation of localized bulging in an inflated hyperelastic tube with fixed ends. {\it Int. J. Appl. Mech.} 12(10), 2050118 (2020). \url{https://doi.org/10.1142/S1758825120501185}
	
	\bibitem{WGZLF} Wang, S., Guo, Z., Zhou, L., Li, L.,  Fu, Y.B.: An experimental study of localized bulging in inflated cylindrical tubes guided by newly emerged analytical results. {\it J. Mech. Phys. Solids} 124, 536--554 (2019). \url{https://doi.org/10.1016/j.jmps.2018.11.011}
	
	\bibitem{VD} Varatharajan, N., DasGupta, A.: Study of bifurcation in a pressurized hyperelastic membrane tube enclosed by a soft substrate. {\it Int. J. Non-linear Mech.}
	95, 233--241 (2017). \url{https://doi.org/10.1016/j.ijnonlinmec.2017.05.004}
	
	\bibitem{WAF} Wang, J., Althobaiti, A.,  Fu, Y. B.: Localized bulging of rotating elastic cylinders and tubes. {\it J. Mech. Mater. Struct.} 12(4), 545--561 (2017). \url{http://dx.doi.org/10.2140/jomms.2017.12.545}
	
	\bibitem{WF} Wang, J.,  Fu, Y.B. (2018).: Effect of double-fibre reinforcement on localized bulging of an inflated cylindrical tube of arbitrary thickness. {\it J. Eng. Math.} 109(1), 21--30. \url{https://doi.org/10.1007/s10665-017-9899-5}
	
	\bibitem{LYAX} Liu, Y., Ye, Y., Althobaiti, A., Xie, Y. X.: Prevention of localized bulging in an inflated bilayer tube. {\it Int. J. Mech. Sci.} 153, 359--368 (2019). \url{https://doi.org/10.1016/j.ijmecsci.2019.01.028}
	
	\bibitem{YLAX} Ye, Y., Liu, Y., Althobaiti, A.,  Xie, Y. X.: Localized bulging in an inflated bilayer tube of arbitrary thickness: Effects of the stiffness ratio and constitutive model. {\it Int. J. Solids Struct.} 176, 173--184 (2019). \url{https://doi.org/10.1016/j.ijsolstr.2019.06.009}
	
	\bibitem{HHS} Hejazi, M., Hsiang, Y.,  Srikantha Phani, A.: Fate of a bulge in an inflated hyperelastic tube: theory and experiment. {\it Proc. Roy. Soc. A} 477(2247), 20200837 (2021). \url{https://doi.org/10.1098/rspa.2020.0837}
	
	\bibitem{FJG} Fu, Y.B., Jin, L., Goriely, A.: Necking, beading, and bulging in soft elastic cylinders. {\it J. Mech. Phys. Solids} 147, 104250 (2021). \url{https://doi.org/10.1016/j.jmps.2020.104250}
	
	\bibitem{EFa} Emery, D.,  Fu, Y.B.: localized bifurcation in soft cylindrical tubes under axial stretching and surface tension. {\it Int. J. Solids Struct.} 219, 23--33 (2021). \url{https://doi.org/10.1016/j.ijsolstr.2021.02.007}
	
	\bibitem{EFb} Emery, D.,  Fu, Y.B.: Post-bifurcation behaviour of elasto-capillary necking and bulging in soft tubes. {\it  Proc. R. Soc.} A 477, 20210311 (2021). \url{https://doi.org/10.1098/rspa.2021.0311}
	
	
	
	\bibitem{WJF2022} Wang, M., Jin, L.S., Fu, Y.B.: Axi-symmetric necking versus Treloar-Kearsley instability in a hyperelastic sheet under equibiaxial stretching.  {\it Math. Mech. Solids}, February 2022. doi:10.1177/10812865211072897. \url{https://doi.org/10.1177/10812865211072897}
	
	\bibitem{O} Ogden, R.W.:  {\it Non-linear elastic deformations.} Ellis Horwood, New York (1984).
	
	\bibitem{HO} Haughton, D.M.,  Ogden, R.W.: Bifurcation of inflated circular cylinders of elastic material under axial loading - I. Exact theory for thick-walled tubes. {\it J. Mech. Phys. Solids} 27(5-6), 489--512 (1979). \url{https://doi.org/10.1016/0022-5096(79)90001-2}
	
	\bibitem{K} Kirchg\"{a}ssner, K.: Wave-solutions of reversible systems and applications. {\it J. Diff. Eq.}, 45(1), 113--127 (1982). \url{https://doi.org/10.1016/0022-0396(82)90058-4}
	
	\bibitem{M} Mielke, A.:  {\it Hamiltonian and Lagrangian flows on center manifolds, with applications to elliptic variational problems}. Springer-Verlag, Berlin
	Lecture Notes in Mathematics (1991).
	
	\bibitem{HI} Haragus, M., Iooss, G.: {\it Local bifurcations, center manifolds, and normal forms in infinite-dimensional dynamical systems}. Springer, London (2011).
	
	\bibitem{WF2020} Wang, M.,  Fu, Y.B.: Necking of a hyperelastic solid cylinder under axial stretching: Evaluation of the infinite-length approximation. {\it Int. J. Eng. Sci.}, 159, 103432 (2021). \url{https://doi.org/10.1016/j.ijengsci.2020.103432}
	
	
\end{thebibliography}
\end{document}